# Rise of Generative Artificial Intelligence in Science

Liangping Ding [a], Cornelia Lawson [a,] and Philip Shapira [a,b,*]

*liangping.ding@manchester.ac.uk*
**0000-0001-6832-4114**
*cornelia.lawson@manchester.ac.uk*
**0000-0002-1262-5142**
\* *pshapira@manchester.ac.uk*
**0000-0003-2488-5985**

[a] Manchester Institute of Innovation Research, Alliance Manchester Business School, The University of Manchester, UK.
[b] School of Public Policy, Georgia Institute of Technology, USA.

**Abstract**

Generative Artificial Intelligence (GenAI, generative AI) has rapidly become available as a tool in scientific research. To explore the use of generative AI in science, we conduct an empirical analysis using OpenAlex. Analyzing GenAI publications and other AI publications from 2017 to 2023, we profile growth patterns, the diffusion of GenAI publications across fields of study, and the geographical spread of scientific research on generative AI. We also investigate team size and international collaborations to explore whether GenAI, as an emerging scientific research area, shows different collaboration patterns compared to other AI technologies. The results indicate that generative AI has experienced rapid growth and increasing presence in scientific publications. The use of GenAI now extends beyond computer science to other scientific research domains. Over the study period, U.S. researchers contributed nearly two-fifths of global GenAI publications. The U.S. is followed by China, with several small and medium-sized advanced economies demonstrating relatively high levels of GenAI deployment in their research publications. Although scientific research overall is becoming increasingly specialized and collaborative, our results suggest that GenAI research groups tend to have slightly smaller team sizes than found in other AI fields. Furthermore, notwithstanding recent geopolitical tensions, GenAI research continues to exhibit levels of international collaboration comparable to other AI technologies.

# 1. Introduction

Artificial intelligence (AI) promises to revolutionize science and technology, sparking widespread discussions about its implications, including on ethics (OECD, 2023), reproducibility (Haibe-Kains et al., 2020), labor market shifts (Septiandri et al., 2024), productivity (Filippucci et al., 2024), and creativity (Mukherjee & Chang, 2024). These discussions have been heightened by the recent and rapid rise of Generative Artificial Intelligence (GenAI) – a type of AI that uses machine learning to dynamically generate new content from large amounts of training data (Lorenz et al., 2023; Stryker & Scapicchio, 2024). GenAI (which encompasses large language models such as ChatGPT) has garnered sizable attention within both the scientific community and broader public spheres (Haque et al., 2022), through the potential to accelerate the pace of science, boost the discovery of new drugs and materials, transform experimental design and hypothesis formulation, and disrupt scholarly publishing (Vert, 2023; Liu et al., 2023; Charness et al., 2023; Lund et al., 2023; Harries et al., 2024). A 2023 Nature survey revealed a substantial uptake of generative AI tools among researchers, with a significant proportion leveraging these technologies to brainstorm ideas and conduct research (Van Noorden & Perkel, 2023).

However, notwithstanding its rapid spread, the deployment of generative AI remains fraught with ethical and epistemic challenges (Blau et al., 2024). Generative models are prone to produce erroneous or fictitious outputs (or "hallucinations") (Jin et al., 2023). Moreover, the non-transparent nature of proprietary generative AI systems, exemplified by ChatGPT's closed-source architecture, raises fundamental questions regarding intellectual property rights and ethical responsibilities in scientific research (Liverpool, 2023; Lund et al., 2023).

In these debates about GenAI's potential benefits and drawbacks, we suggest that probing the extent of its use is important in helping to avoid seriously overrating or underrating its applications in science. In this paper we investigate three critical research questions to better understand the diffusion and impact of GenAI in science. First, we explore whether GenAI technology is indeed diffusing into the sciences and, if so, what the current interdisciplinary diffusion landscape looks like. Recent studies have observed the rapid adoption of GenAI tools in various scientific disciplines, reflecting AI's growing influence on research methodologies and knowledge production (Jiang et al., 2023). Understanding how GenAI is being adopted across different scientific fields will provide insights into its role in shaping contemporary scientific practices.

Second, we investigate how GenAI influences collaboration patterns in scientific research. It is reported that GenAI is helping some researchers to write code, review literature, and prepare presentations and manuscripts (Owens, 2023), with Noy & Zhang (2023) finding in a controlled experiment that college-educated professionals using ChatGPT saw substantial productivity gains across their various writing tasks. Conceivably, if GenAI leads to increased productivity across the range of tasks undertaken by researchers, this *might* lead to reduced team sizes. Indeed, concerns have been raised about the potential of GenAI to replace jobs, including those traditionally held by human researchers (Kim, 2023). On the other hand, as scientific research becomes increasingly specialized and collaborative (Jones, 2009), with large teams often required to tackle complex problems across various disciplines (Venturini et al., 2024), it is also conceivable that GenAI *might* lead to expanded team sizes. In short, the impact of GenAI on team size and composition is still unclear. Understanding whether GenAI might reduce the need for large, diverse teams by automating certain tasks and roles, or conversely, necessitate even larger collaborations, is important for anticipating future research dynamics and human resource implications.

Our third question concerns the potential of GenAI to influence international collaborations. International collaboration is often associated with high-quality research outcomes and increased citation rates (Wang et al., 2024). However, geopolitical standoffs between major research nations, such as China and the United States, Russia's war in Ukraine, and other international tensions, pose new challenges to global scientific collaboration (Jia et al., 2024; Min et al., 2023; Shih et al., 2024). In this context, examining how GenAI (which has risen contemporaneously with recent global tensions) might either bridge or exacerbate these divides is particularly timely. The intersection of GenAI with international collaboration offers a rich avenue for understanding the broader implications of GenAI in a rapidly changing global landscape.

To address these questions, this paper presents an exploratory bibliometric analysis of the rise of GenAI in scientific research. Using OpenAlex, which provides comprehensive scientific publication metadata (Priem et al., 2022), we analyze selected populations of publications to investigate the characteristics of GenAI compared to other AI technologies. In this respect, we seek to explore the additional impacts of GenAI compared with developments already underway through the deployment of earlier and more well-established AI technologies. Specifically, we examine the disciplinary and geographic diffusion of GenAI, as well as its impact on team size and international collaboration. Updated findings on these topics will contribute to the ongoing discourse on GenAI's role in science, providing insights that are relevant for research initiatives and science policy.

## 2. Data and Methodology

*2.1 Data collection*

We construct a bibliometric sample containing both GenAI papers and other artificial intelligence (Other-AI) papers to investigate the rise of GenAI in science compared with other AI technologies. We employ OpenAlex as our primary data source, leveraging its extensive repository of over 250 million scientific works, encompassing journal articles, conference proceedings, and more. The dataset is sourced from the January 2024 snapshot of OpenAlex and our study retains papers published before 2024. We use a two-step process.

In the first step, to identify GenAI related papers in OpenAlex, we build a list of search keywords, informed by prior bibliometric studies. Kanbach et al. (2024) composed a bibliometric definition for generative AI. Although there are multiple GenAI models, they demonstrated that there is a clear tendency for most of these publications to use ChatGPT as the most prominent application of GenAI. In contrast, Mariani & Dwivedi (2024) deployed a list of model-name keywords to cover GenAI, including generative AI models such as "Runway" and "Bard". However, we find that these keywords add noise into the retrieved results where GenAI models are named after commonly found terms that are not unique to GenAI. We thus proceeded by formulating a list of terms and phrases specific to generative AI, with ChatGPT serving as a representative model, and constructed PySpark SQL queries targeting the title and abstract fields.[1] (See Table 1.) We further designed a Boolean search statement to filter SQL records based on specific string matches, employing the 'LIKE' typed operator along with wildcard characters '%' to account for variations in terms. Both 'LIKE' and 'ILIKE' operators are utilized to differentiate between case-sensitive and case-insensitive terms.

<Table 1 about here>

The publications returned by the search terms are not exclusive; for example, a publication containing both 'chatgpt and 'Generative AI' would be captured by both search terms. We subsequently de-duplicate the records to address this. Additionally, we note that the acronym 'GPT' is also used elsewhere in science other than AI research. For example, GPT can refer to 'Glutamate Pyruvate Transaminase' in biotechnology research. To refine the dataset, we developed a preliminary set of GPT full-expansion terms through an intensive iterative process of manually checking the retrieved results to address the ambiguity of the abbreviation 'GPT'. Table 2 presents the complete list of expansions for the acronym GPT that are excluded from our study. We further exclude irrelevant terms using the Boolean operator 'NOT'.

---

[1] There are other GenAI models such as LlaMA (introduced in February 2023). However, given the 2017-2023 time span of our study, we decided to focus on the forerunner and most prominent GenAI models, i.e. GPT series models. Adding LlaMA into the search terms introduces non-relevant publications as llama is a common word found in zoology, nature conservation, environmental science, and other disciplines. We also observed that "AIGC" (Artificial Intelligence Generated Content), a term used in China to describe generative AI, appears in fewer than 500 publications within our study's timeframe. Due to its limited prevalence, we did not include this term in our analysis.

<Table 2 about here>

In the second step, for Other-AI papers, we also used a keyword approach, beginning first with an approach to identify AI. We recognize that with its diverse interdisciplinary roots, varied development paths, and rapid recent growth (introducing new models and terms), 'artificial intelligence' is a challenging concept to define and quantify (Baruffaldi et al., 2020). Prior studies have constructed AI bibliometric search terms (Liu et al., 2021; Van Noorden & Perkel, 2023), although no consensus has been reached in the bibliometric community on defining AI. Considering the ongoing development of AI, we follow the AI search terms from the recent study by Van Noorden & Perkel (2023, Supplementary Information, AI survey methodology). Specifically, we search for papers in OpenAlex spanning 2017-2023, with titles or abstracts containing the following search terms: 'machine learning'; 'neural net*', 'deep learning', 'random forest', 'support vector machine', 'artificial intelligence', 'dimensionality reduction', 'gaussian process', 'naïve bayes', 'large language model', 'llm*', 'gaussian mixture model', and 'ensemble methods'. We exclude 'chatgpt' in the search, and then further exclude all the GenAI papers identified in the first step. This allows us to build a corpus of Other-AI papers based on the OpenAlex unique publication identifier, ensuring that there is no overlapping between these two corpora.

*2.2 Data preprocessing*

To deal with duplications, missing metadata and non-research publication records in our retrieved results, we implement a series of data preprocessing procedures, to arrive at a refined dataset for subsequent analysis. These steps are listed in Appendix 1. The refined dataset comprises 14,417 GenAI entries and 1,422,683 Other-AI entries after data preprocessing.

To explore the disciplinary diffusion of GenAI, we associate each of these papers with their subject to explore the interdisciplinary diffusion of GenAI technology. We follow the methodology introduced by Klebel & Ross-Hellauer (2023) to leverage the concepts entity provided within OpenAlex as the subject. We use the 19 root-level concepts from OpenAlex, which comprise (in alphabetical order) Art, Biology, Business, Chemistry, Computer science, Economics, Engineering, Environmental science, Geography, Geology, History, Materials science, Mathematics, Medicine, Philosophy, Physics, Political science, Psychology, and Sociology. Additionally, OpenAlex assigns a score to indicate the strength of association between a publication and a given concept. Concepts with a score of 0 were filtered out from our analysis.

To explore the geographical diffusion of GenAI, we focus on the authorship country distribution, but this analysis was complicated by the high proportion of missing institutional metadata in OpenAlex. Zhang et al. (2024) identified a prevalent issue of missing institution data in more than 60% of journal articles within the OpenAlex database. In OpenAlex, a publication is associated with authorships, each potentially linked to one or more institutions, delineated by five primary fields: institution, institution name, Research Organization Registry (ROR), country code, and institution type. In our dataset, among the 14,417 GenAI publications obtained from OpenAlex, 8,653 (60.0%) publications have missing institutions. Among the 1,422,683 AI publications, 465,721 publications (32.7%) have missing institutions. The term "missing institution" refers to the absence of any of the metadata including institution ID, institution name and institution country code.

The substantial proportion of missing institutional metadata in the OpenAlex database, particularly for GenAI publications, presented significant challenges to data completeness. To address this issue, we conducted additional preprocessing based on the refined disciplinary diffusion dataset. This involved filtering publications to retain only those with complete institutions for both GenAI and other AI publications and supplementing missing institutional data for GenAI publications by scraping institutional information from the web. Given the high rate of missing institutional metadata for GenAI publications, it became necessary to use two distinct datasets for this study: one dedicated to disciplinary diffusion and another focused on geographical diffusion and empirical analysis to investigate the impact of GenAI on collaboration intensity and international collaboration.

To supplement the incomplete institutional data of GenAI publications, particularly for ArXiv preprints where many papers were missing institutional metadata, we employed two methods to supplement our data set. First, institutional metadata at the paper level was retrieved from the OpenAlex webpage. Second, for ArXiv preprints, PDF documents were processed to extract institution information. Through plain-text conversion, we identified the page containing author and institutional details. Institution names and corresponding countries were extracted using GPT-3.5. The prompt used is illustrated in Fig. 1. To ensure consistency, the extracted institutions were normalized against the OpenAlex affiliation database. This normalization process involved constructing a term-frequency inverse-document-frequency (TF-IDF) vector space model and calculating cosine similarity scores between GPT-3.5-extracted institutions and OpenAlex database entries. The most similar matches, determined by cosine similarity scores, were selected as the normalized institutional names. After correcting for missing institutional data, we had a dataset of 967,640 publications for geographical diffusion analysis and empirical analysis. This included 10,678 GenAI-related publications and 956,962 other-AI related publications.

<Fig. 1 about here>

*2.3 Empirical Model*

To explore the impact of GenAI on collaborations, we construct measures of GenAI, which is a dummy variable coded 1 if a publication is related to GenAI and 0 for Other-AI. The collaboration intensity (*CI*) for each publication is calculated as the number of authors listed on each publication. International collaboration intensity (*ICI*) is calculated as the number of countries represented by the authors' affiliations.

We estimate two models to explore the collaboration intensity and international collaboration intensity of GenAI related research. Both measures are count variables and we therefore estimate Poisson models as follows:

$$\ln(CI_i^{(1)}) = \beta_1 GenAI_i^{(1)} + \sum_{t=1}^{T-1} \gamma_t \text{Year}_t + \sum_{j=1}^{J-1} \delta_i \text{Subject}_i + \epsilon_i^{(1)} \qquad (1)$$

$$\ln(ICI_i^{(2)}) = \beta_2 GenAI_i^{(2)} + \beta_3 CI_i^{(2)} + \sum_{t=1}^{T-1} \gamma_t \text{Year}_t + \sum_{j=1}^{J-1} \delta_i \text{Subject}_i + \epsilon_i^{(2)} \qquad (2)$$

Where *GenAI* indicates whether the publication is concerned with generative AI. *CI* is the number of authors for each publication $i$. *ICI* is the number of countries for each publication $i$. As usual, $\epsilon_i$ is the error term while $\gamma_t$ and $\delta_i$ stand for year fixed effects and subject effects respectively to account for year-specific influences (ranging from 2017 to 2023) and disciplinary differences (with the subject score assigned by using the confidence score of the 19 root-level concepts in OpenAlex).

**3. Results**

*3.1 Publication trends of GenAI across disciplines*

Fig. 2 illustrates the annual trend in overall scientific publications as well as the disciplinary-specific publication trends. We can see that while Other-AI publication output grew significantly from 2017-2023, there was an even faster growth in GenAI publications, with a surge from 2022 (See yearly growth rate in Appendix Fig. S1). This exponential growth pattern reflects the recent surge of engagement with generative AI. Fig. 2 also illustrates the distribution of generative AI publications across scientific domains, providing a visual representation of the diffusion of generative AI across all scientific domains. GenAI publications have boomed in computer science, although Other-AI research in computer science continues to be far greater (10x) in scale. A similar pattern is seen in the growth and scale relationships of Other-AI and GenAI in medicine. In fields such as art, sociology, and psychology, GenAI has captured a relatively greater share compared with Other-AI. Possibly this reflects ease of entry (e.g. for GenAI-generated visualizations), debate about its societal implications, and attention to implications for education and cognitive development (e.g. in psychology). Overall, the wide-ranging diffusion of generative AI publications across diverse scientific domains highlights its interdisciplinary nature and its potential to influence change in approaches, practices, and methods across multiple research fields.

<center><Fig. 2 about here></center>

*3.2 Geographical distribution of Generative AI*

Fig. 3 presents the publication output of the top 20 most productive countries in GenAI and Other-AI research from 2017 to 2023[2]. Each bar shows the percentage contribution of a given country/region to global GenAI and Other-AI publications. The United States and China are the two dominant performers of AI research, collectively contributing a substantial proportion of global publications in both GenAI and Other-AI domains. The United States leads GenAI research with 39% of global output, while China dominates in Other-AI research with 27.9% of publications. This contrast between GenAI and Other-AI publications in the United States and China highlights that high publication outputs in broader AI research, as exemplified by China's leadership in Other-AI, does not necessarily translate to a leading role in GenAI research. In the emerging field of GenAI, the United States has decisively outpaced other nations in its publication output share through to 2023. This finding underscores the pivotal role of the United States in driving the rapid advancement of GenAI technologies.

<center><Fig. 3 about here></center>

Other countries lag significantly behind the US and China in terms of their shares of world GenAI and Other-AI publication outputs. Nations like the UK, India, Germany, Canada, Australia, Japan, and Italy have lower shares of global output, with contributions typically ranging between 2% and 8% in GenAI and Other-AI fields. However, when considering the ratio of 2017-2023 world public output shares in GenAI to those in Other-AI, Hong Kong and Singapore stand out, with ratios of 2.6 and 2.2, respectively. The US ratio by this measure is 1.8, followed by Switzerland (1.5), and Australia and the UK (each with 1.4). In other words, a set of advanced economies (including smaller ones as well as the much larger US) have seen relatively higher shifts in their AI publication outputs towards GenAI. Mainland China (with a ratio of 0.5) has not shifted so rapidly. This diversification of research output across different nations highlights the global landscape of AI research and the evolving dynamics between GenAI and Other-AI fields. It remains to be seen how the observed patterns reflect broader trends towards specialization in AI research and how these patterns might evolve in the face of new technological opportunities in the AI landscape.

---

[2] The ISO two-letter country code is used to represent the country.

*3.3 Regression Results*

To investigate the impact of GenAI on collaborations, we estimate the two equations presented in section 2.3.

Table 3 provides descriptive statistics for the key variables. The skewness and kurtosis values of the variables reveal that *CI* is highly skewed (>300) and exhibits excess kurtosis (>18K), indicating the presence of numerous extreme outliers. To address this, we remove the outliers by applying a 99th percentile threshold, thus reducing the number of observations to 958,343. On average, each paper has 4.27 authors and involves 1.30 countries. Regarding the distribution of subjects, the average subject confidence scores indicate that most publications are in Computer Science (mean confidence score of 49%), followed by Medicine (10%), Mathematics (5%), Engineering (4.5%), and Biology (3.5%).

<center><Table 3 about here></center>

The baseline regression results are presented in Table 4. The dependent variable Collaboration Intensity (*CI*) is shown in column (1) and International Collaboration Intensity (*ICI*) is shown in column (2). We estimate a Poisson model with controls for year and subject fixed effects, reporting marginal effects with robust standard errors. The results are robust to OLS estimates using logged *CI/ICI*. Fig. 4 shows the estimated margins of *CI* and *ICI* for GenAI related and Other-AI related publications.

<center><Table 4 about here></center>

The results show that in column 1 (*CI*), the marginal effect of GenAI is negative and significant, suggesting a negative correlation between GenAI and collaboration intensity, compared to Other-AI. Fig. 4 graphically shows that GenAI related publications tend to have smaller team sizes compared to other-AI related publications (4.18 vs. 4.28, p< .001). This finding raises important questions about the nature of collaboration in GenAI research. One possible interpretation is that the adoption of GenAI tools might be enhancing researchers' productivity, reducing somewhat the need for large, multidisciplinary teams typically associated with Other-AI research and could thus point to a more streamlined organizational structure in GenAI-driven research environments. Alternatively, the relatively easier accessibility of GenAI through APIs may lower the barriers to entry, allowing researchers with limited advanced AI expertise to engage in GenAI-related studies. In this case, the reduction in collaboration intensity could reflect that fewer specialists are required to conduct the research compared to Other-AI. However, caution must be used when interpreting these results, as the observed effect is small, and teams remain relatively large in both cases (>4 authors).

<Fig. 4 about here>

By contrast, in column (2) Table 4, we can see that GenAI is positively related to international collaboration intensity, indicating that GenAI research involves more international collaboration compared to Other-AI research. This is presented graphically in Fig 4 (1.35 vs. 1.30, p< .001). This suggests that while GenAI research may foster smaller team sizes, it may simultaneously encourage a more globally connected network, possibly due to the collaborative nature of cutting-edge GenAI applications and the shared international interest in GenAI technologies. Again, the differences are small, and the majority of publications do not involve international collaboration for both GenAI and Other-AI publications (median= 1). Regardless, this divergence between team sizes and international collaboration warrants further investigation, particularly in exploring whether it reflects differing research motivations, the accessibility of GenAI tools, or broader geopolitical trends in AI research.

## 4. Discussion and Implications

Our initial exploratory analysis reveals that the relevance of GenAI in scientific research has expanded well beyond its origins in computer science. While early developments in GenAI were predominantly concentrated within the computer science field, we now observe the broader diffusion of these technologies across a diverse range of scientific disciplines. This includes fields such as medicine, chemistry, geography, and even sociology, where researchers are increasingly leveraging GenAI for tasks such as data generation, predictive modeling, and hypothesis testing. This cross-disciplinary adoption suggests that GenAI is emerging as a general purpose technology (Eloundou et al., 2024) and is becoming a tool used not just within its foundational domain, but as an innovation for enhancing research methodologies, accelerating discovery, and addressing complex scientific challenges in a variety of fields.

We find that the US has moved more rapidly into using GenAI in science publication fields than China (through to 2023). The US research community appears to be contributing to and exploiting US leadership in the innovation and implementation of GenAI, aided by the country's dynamic tech ecosystem, high levels of R&D investment, massive funding for GenAI model development both by large US tech companies and venture capital-sponsored new entrants, and AI policies that prioritize cutting-edge research and development (Fattorini et al., 2024; NSTC, 2023). China's continued high rates of publication in Other-AI reflects its own extensive R&D efforts, deep investments in mainstream AI technologies, national policy frameworks, and the emergence of a co-evolved public-private AI innovation ecosystem (Lundvall & Rikap, 2022; State Council, 2024). Although China was the second largest producer of GenAI publications through to 2023, this is outweighed by its far larger base of publication outputs in Other-AI. China is now accelerating efforts, particularly on the private-sector side, to strengthen the development and use of GenAI, including building its own generative AI models albeit with concerns about open access and governance (Triolo & Schaefer, 2024; Chang, 2024). While the Chinese scientific research community seems to have been relatively slower to date in engaging with GenAI, deployment can be expected to grow going forward. Researchers in numerous other economies are also engaging with GenAI in their AI-related scientific publications, with relatively high levels of deployment seen in several small and medium-sized advanced economies.

Our regression results suggest that research teams focusing on GenAI tend to be slightly smaller compared to those working on other forms of AI, controlling for year-specific and disciplinary differences. Reductions in team size may, in part, reflect increasing productivity enabled by advances in GenAI or its easier accessibility. Our findings partially align with the study by Thu et al. (2022), which examines team sizes in machine learning (ML)-related and ML-unrelated projects. Their research indicates that ML-related projects tend to involve slightly fewer authors compared to ML-unrelated ones, with no significant difference in the number of countries represented. They suggest this might be attributed to the reduced need for physical work in ML-related projects, thereby requiring fewer team members. As GenAI further evolves and diffuses, researchers may benefit from more efficient workflows, automated processes, and enhanced model capabilities. These may help streamline tasks such as data generation, model training, and experimentation, although it is also possible that productivity effects could be dampened by research task amplification, additionality and diversification (Ribeiro et al., 2023). Additionally, despite the trend toward smaller team sizes, we find that GenAI research is slightly more international than for Other-AI. This is noteworthy in that it suggests that the level of international cooperation in GenAI research remains at least on par with, if not somewhat exceeding, the levels found in Other-AI fields notwithstanding the extensive array of geopolitical tensions that have arisen coincident with GenAI's rapid growth in recent years.

We acknowledge that our study has limitations, including those related to definitions and data, although we have sought to be careful in implementing our search approach and cleaning the dataset. We also recognize that there are a variety of ways that scientists are using GenAI in their research publications as well as in research practices not necessarily captured by publication metadata. Our study, while exploratory, does indicate that GenAI is being increasingly adopted by scientists in their research across the multiple fields of science. The net productivity effects of GenAI's increased use in science and how it will influence team sizes and international collaborations – as well as its implications for creativity, reproducibility, ethics, and other concerns – clearly all represent important topics for future research that can build upon these early findings.

**Acknowledgments**
We are grateful for feedback received from Cassidy Sugimoto and John Walsh at the 2024 Atlanta Academy on Science and Innovation Policy.

**Author contributions**
Liangping Ding: Conceptualization; Methodology; Data curation; Formal Analysis; Writing.
Cornelia Lawson and Philip Shapira: Conceptualization; Methodology; Writing; Supervision; Funding acquisition

**Competing interests**
The authors declare no competing interests.

**Funding information**
The authors acknowledge support from the University of Manchester Faculty of Humanities Research Investment Fund.

**Table 1. Generative AI Search Terms**

| Signal Term | Case Sensitive |
|---|---|
| Chatgpt | N |
| Generative AI OR generative AI | Y |
| GPT-3 | Y |
| GPT-4 | Y |
| Generative artificial intelligence | N |
| Generative pretrained transformer | N |
| GPT-2 | Y |
| Generative language model | N |
| GenAI | Y |
| Generative large language model | N |
| OpenAI GPT | N |
| Generative pretrained language model | N |
| GPT-1 | Y |
| OpenAI large language model | N |

Source: Authors elaboration. See discussion in text.

**Table 2. Expansions of the acronym GPT**

| Full expansion of the acronym GPT (Case Insensitive) |
|---|
| Glutamate Pyruvate Transaminase |
| glutamic-pyruvic transaminase |
| N-p-tolyl-D-glucosylamine |
| UDP-GlcNAc:dolichol-P GlcNAc-1-P transferase |
| Goniopora toxin |
| Gradational psychosomatic treatment |
| Gas production technique |
| guinea-pig isolated trachea |
| gemcitabine chemotherapy |
| General Purpose Technology |
| UDP-GlcNAc:dolichyl-phosphateN-acetylglucosamine 1-phosphate transferase |
| General Professional Training |
| Glc6P/phosphate translocator |
| gemcitabine-concurrent proton radiotherapy |

Source: Authors elaboration. These expansions of GPT are subsequently excluded in bibliometric search. See discussion in text.

**Table 3. Descriptive Statistics**

|  | mean | sd | min | max |
|---|---|---|---|---|
| GenAI | 0.011 | 0.104 | 0.000 | 1.000 |
| CI | 4.274 | 2.547 | 1.000 | 16.000 |
| ICI | 1.302 | 0.671 | 1.000 | 15.000 |
| Art | 0.002 | 0.024 | 0.000 | 0.801 |
| Biology | 0.035 | 0.122 | 0.000 | 0.975 |
| Business | 0.014 | 0.075 | 0.000 | 0.840 |
| Chemistry | 0.018 | 0.089 | 0.000 | 0.994 |
| Computer science | 0.491 | 0.288 | 0.000 | 0.956 |
| Economics | 0.007 | 0.047 | 0.000 | 0.859 |
| Engineering | 0.045 | 0.101 | 0.000 | 0.734 |
| Environmental science | 0.019 | 0.097 | 0.000 | 0.865 |
| Geography | 0.013 | 0.059 | 0.000 | 0.777 |
| Geology | 0.010 | 0.060 | 0.000 | 0.949 |
| History | 0.001 | 0.019 | 0.000 | 0.694 |
| Materials science | 0.020 | 0.099 | 0.000 | 0.967 |
| Mathematics | 0.050 | 0.111 | 0.000 | 0.954 |
| Medicine | 0.098 | 0.227 | 0.000 | 0.997 |
| Philosophy | 0.002 | 0.026 | 0.000 | 0.825 |
| Physics | 0.017 | 0.077 | 0.000 | 0.978 |
| Political science | 0.010 | 0.057 | 0.000 | 0.787 |
| Psychology | 0.029 | 0.105 | 0.000 | 0.912 |
| Sociology | 0.006 | 0.047 | 0.000 | 0.819 |

Source: Analysis of OpenAlex (see discussion in text). N= 958,343

**Table 4. Regression results of GenAI on Collaboration Intensity (CI) and International Collaboration Intensity (*ICI*)**

|            | (1)          | (2)          |
|------------|--------------|--------------|
|            | ***CI***     | ***ICI***    |
|            |              |              |
| GenAI      | -0.100***    | 0.0490***    |
|            | (-3.46)      | (7.59)       |
|            |              |              |
| CI         |              | 0.0749***    |
|            |              | (206.20)     |
|            |              |              |
| Year FE    | Y            | Y            |
| Subject FE | Y            | Y            |
|            |              |              |
| N          | 958,343      | 958,343      |
| ll         | -2063883.0   | -1146085.7   |
| Pseudo R2  | 0.0524       | 0.0119       |

Note: Marginal effects are reported, robust standard errors in parentheses. Results are robust to OLS estimates using logged CI/ICI.
* $p<0.05$, ** $p<0.01$, *** $p<0.001$

**Fig. 1: Prompts used to address for missing institution completeness**

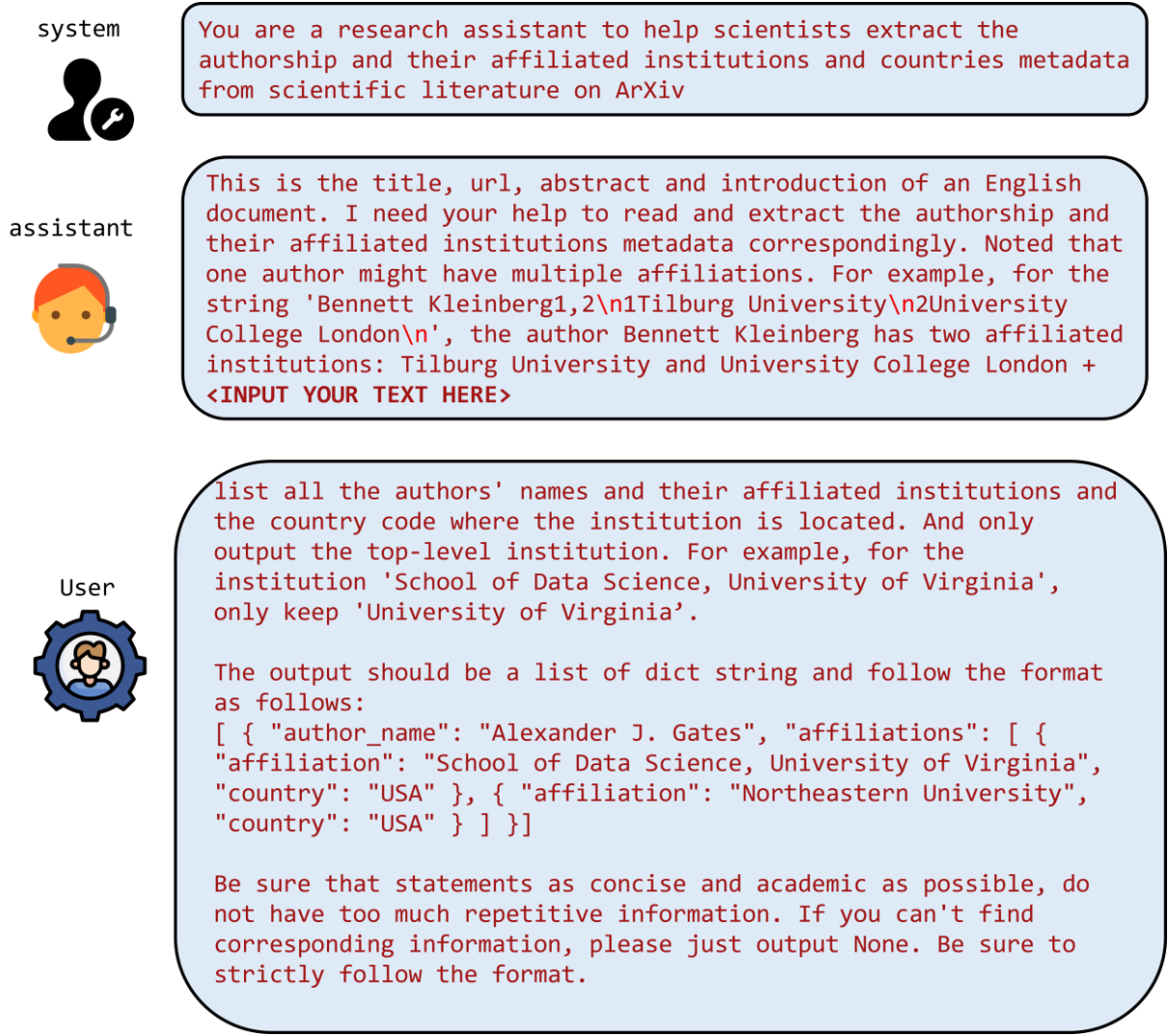

**Fig. 2: Overall and Disciplinary-Specific Publication Trends, 2017–2023**

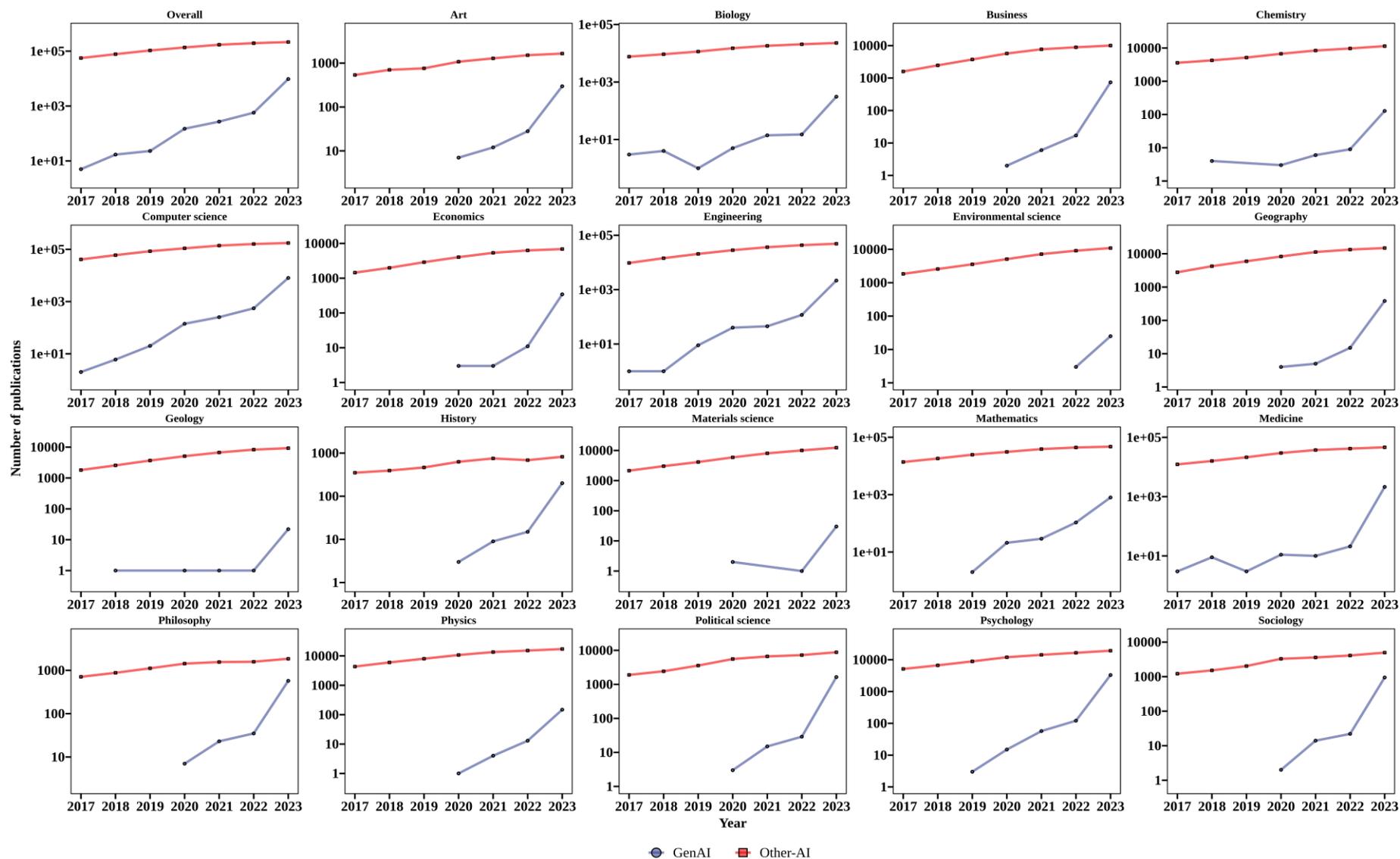

Source: Analysis of OpenAlex publication metadata. Number of records: GenAI = 14,417; Other AI = 1,422,683.

**Fig. 3: Leading Countries for GenAI and Other-AI Publications, 2017-2023**

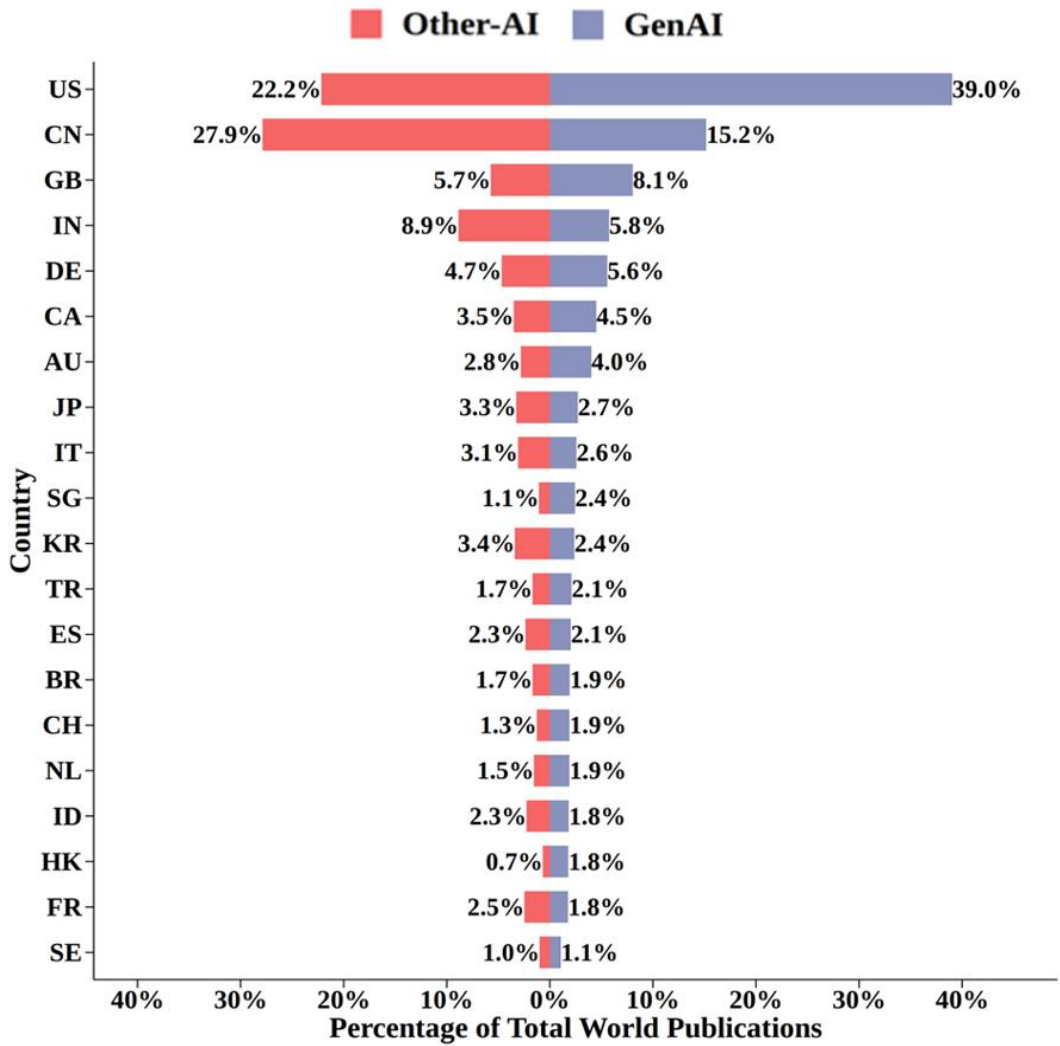

**Fig. 4:** *CI* and *ICI* Estimation by Poisson Regression Controlling for Years and Subjects

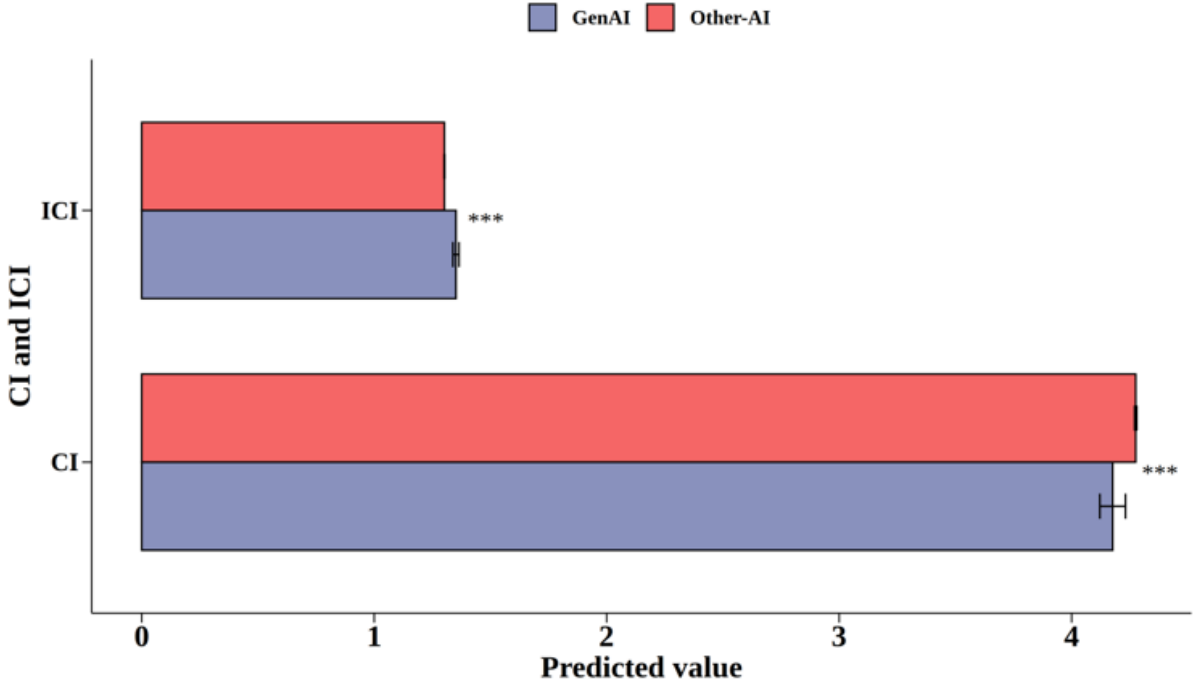

*Note: CI* = Collaboration Intensity; *ICI* = International Collaboration Intensity. The error bars represent the 95% confidence intervals. Two-tailed test: ***p<0.001.

**Appendix 1. Data preprocessing procedures**

The data preprocessing steps include the following:

1. Title Cleaning: Removal of special characters, HTML tags, HTML character entities, and punctuation from entry titles to standardize text format and enhance readability.
2. Filtering by Publication Year: Entries with publication years from 2017 to 2023 were retained.
3. Removal of Entries with Missing DOIs: Elimination of entries lacking Digital Object Identifiers (DOIs) from the dataset.
4. Filtering out Entries with Missing Titles: Removal of entries lacking titles from the dataset.
5. Filtering out Comments and Letters: Removal of entries containing terms such as 'Comment' or 'Correspondence' in the title, indicative of comments or letters.
6. Filtering by Work Type: Selective retention of entries categorized as 'article' or 'book-chapter', including journal-article, proceedings-article, and posted-content, while other work types were filtered out.
7. Deduplication based on Title: Identification of entries with identical titles and retention of only unique entries, employing criteria such as the presence of DOI, source type rank ('journal', 'conference', 'repository', 'book series', 'ebook platform'), version rank ('publishedVersion', 'submittedVersion', 'acceptedVersion'), and earliest publication year.

**Fig. S1. Yearly Growth Rate of Publications: Overall and Across Specific Discipline**

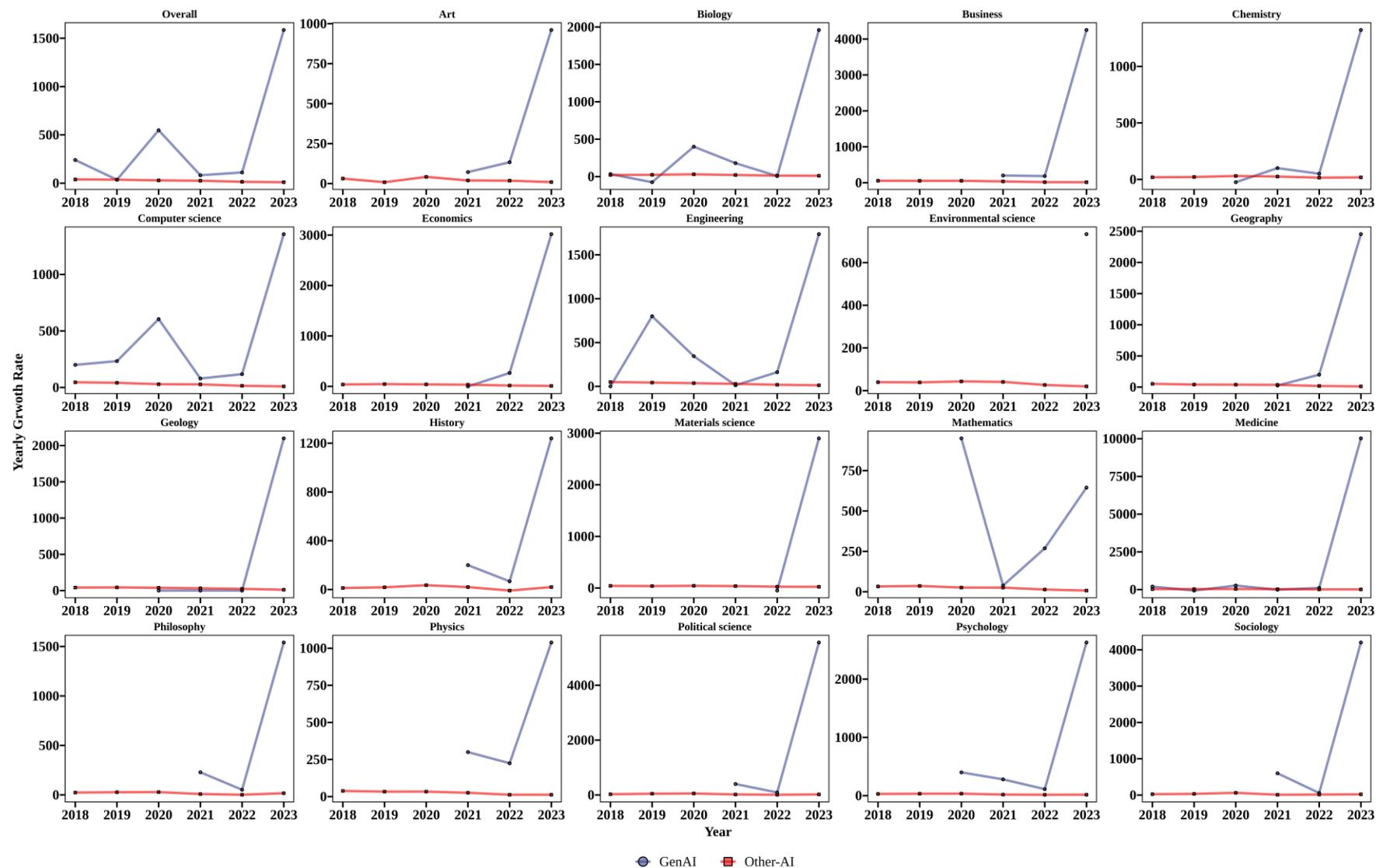

Source: Analysis of OpenAlex publication metadata. Number of records: GenAI = 14,417; Other AI = 1,422,683.